\documentclass[aps,amssymb,twocolumn,nobalancelastpage]{revtex4-2}
\usepackage{graphicx}
\graphicspath{ {Figures/} }
\usepackage{lpic}
\usepackage{epsfig}
\usepackage{float}
\usepackage[colorinlistoftodos, textwidth=\marginparwidth]{todonotes}
\usepackage{hyperref}
\usepackage{xcolor}
\usepackage{amsmath}
\usepackage{amsfonts}

\renewcommand{\vec}[1]{\mathbf{#1}}

\draft 

\begin{document}


\title{A toy model for viscous liquid dynamics} 



\author{Filip Samuelsen}
\email[]{fisa@ruc.dk}
\affiliation{``Glass and Time'', IMFUFA, Department of Science and Environment, Roskilde University, P.O. Box 260, DK-4000 Roskilde, Denmark}

\author{Lorenzo Costigliola} 
\email[]{lorenzoc@ruc.dk}
\affiliation{``Glass and Time'', IMFUFA, Department of Science and Environment, Roskilde University, P.O. Box 260, DK-4000 Roskilde, Denmark}
    
\author{Thomas B. Schr{\o}der}	
\email[]{tbs@ruc.dk}
\affiliation{``Glass and Time'', IMFUFA, Department of Science and Environment, Roskilde University, P.O. Box 260, DK-4000 Roskilde, Denmark} 


\date{\today}


\begin{abstract}
A simple model for viscous liquid dynamics is introduced. 
Consider the surface of the union of hyper-spheres centered at random positions inside a hypercube with periodic boundary conditions. It is argued and demonstrated by numerical simulations that at high dimensions geodetic flows on this surface is a good model for viscous liquid dynamics. 
It is shown that this simple model exhibits viscous dynamics for densities above the percolation threshold in $8$, $12$ and $16$ dimensions. Thus the slowing down of the dynamics, measured by the mean-squared displacement, extends to several orders of magnitude similarly to what is observed in other models for viscous dynamics. Furthermore, the shape of the mean-squared displacement is to a very good approximation the same as for the standard model in simulations of viscous liquids: the Kob-Andersen binary Lennard Jones mixture.
\end{abstract}

\pacs{}

\maketitle 

\section{Introduction}
Simple models have been essential building blocks in the development of physics.
One of the best examples in the glass and liquid community is the hard sphere model.
Even if this model is not able to capture all important aspects of liquid behavior, as for example the 
gas-liquid phase transition, it has been of fundamental importance in the theoretical development due 
to its mathematical simplicity. 
For this model it is possible, for example, to find a simple equation of state \cite{Carnahan1969}, 
to roughly predict the pair correlation function \cite{HansenMcDonald},
and to predict the existence of a dynamical glass transition \cite{Charbonneau2017}.
The development of computer simulations during the last decades has made it possible to study in detail
a variety of different models, getting insight on their differences and similarities. 
In this work a new model for viscous dynamics is introduced.
The reason for introducing the new models is that they can provide different ways of looking at
well know problems and possibly guide in the direction of solving open questions in physics.
One remarkable and recent example is the poly-disperse model introduced by Berthier \emph{et al}\cite{SwapMC}, 
which allowed for simulations to access temperatures regimes deeply below the glass transition 
Another example could be a recent model for soft colloids which includes internal elasticity \cite{Gnan2019}. 
Many newly proposed models are less simple but allows one to explore new regions in the phase diagram or 
give a more accurate description of a physical phenomenon.
The model introduced in this work is in many ways simpler than present models studied in the viscous dynamics literature, 
but at the same time it reproduces the main features of interest.
This model can be thought to be more a generic model of the potential energy landscape \cite{Stillinger1984,Debenedetti1999} than of a specific physical system, similarly to what was previously done e.g. in Ref. \cite{Pedersen2006}.



\section{The model surface}\label{sec:model}

\begin{figure}
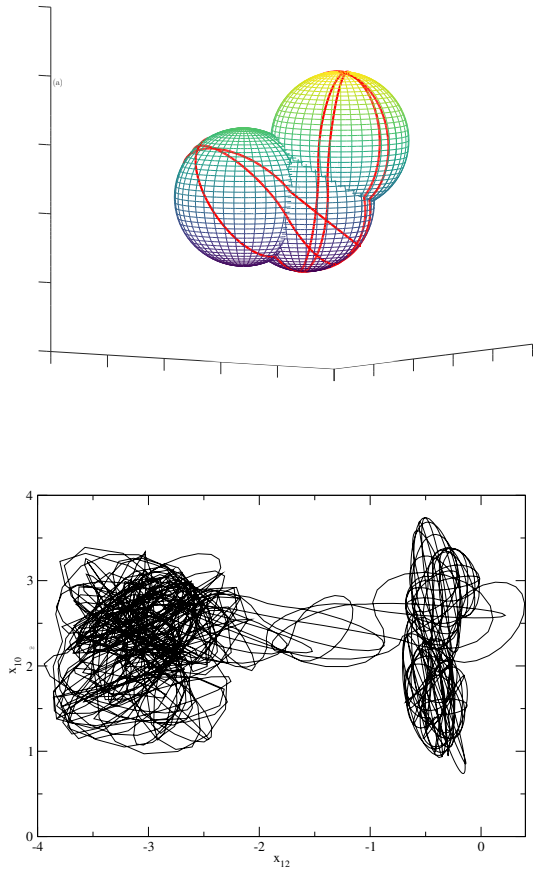

  \centering
  \resizebox{0.5\textwidth}{!}{
  \begin{tabular}{l}
  \begin{lpic}[]{Figures/intersection-inc(,40cm)}\lbl[t]{30,110;\huge{(a)}}\end{lpic} \\
  \begin{lpic}{Figures/flow_event(,40cm)}\lbl[]{30,110;(b)} \end{lpic} 
  \end{tabular}}
  \caption{\label{fig:themodel3d}
  (a) A small $3$-dimensional example of the model proposed in this work. The moving particle will follow geodetics on the surface defined by the union of the spheres.
  Here one can see that the physical intuition of geodesics going straight locally is preserved over the intersections.
  (b) An example of a flow event from a simulation in 16 dimensions, with the trajectory  projected onto a plane. The system is "rattling" in a region for a certain amount of time and then suddenly jumps to a neighboring region, where it again becomes localized. Qualitatively, this is similar to what happens at low temperatures in viscous model liquids, see e.g. \cite{Schroder2000}.
}
\end{figure}

The motivation for the new model is based on 
the so-called NVU dynamics \cite{NVU1,NVU2, quasiuniNVU}. If the positions of all the $N$ particles in a system is represented by a vector $\bf{R}$ 
(which lives in a $D=3N$ Euclidean space), the constant potential energy hyper-surface of interest is $U({\bf R})=U_0$,
where $U_0$ is the average value of the potential energy of the system at some given state point. The NVU dynamics is defined as geodetic flows on this (hyper) surface. Interpreting time as being proportional to the curve length, one finds that the NVU dynamics gives the same particle dynamics, e.g., mean-square displacement and intermediate scattering function, as the traditional NVE and NVT molecular dynamics\cite{NVU1,NVU2}.


The idea is now to make a generic model of the $U({\bf R})=U_0$ surface, and then perform geodetic flows on it. In general the surface $U({\bf R})=U_0$ is a very complicated high dimensional object. Thus to arrive at a simple model, we will make a number of simplifications as described in the following. 

In the viscous regime, the dynamics is well-separated\cite{Schroder2000} into oscillations around the local minima of $U(\bf{R})$ (inherent states), and transitions between these, as originally suggested by Goldstein \cite{Goldstein1969}. It is well known, that close to the inherent states, the potential energy surface is well approximated by a harmonic potential, with a distribution of curvatures. If we know focus first on the $U(\bf{R})$$=U_0$ surface in the vicinity of a single inherent state, this will then be a hyper-ellipsoid. We make the simplification to ignore the distribution of curvatures, and use hyper-spheres instead of hyper-ellipsoids.

Pairs of nearby inherent states can be connected by a potential energy barrier. If $U_0$ is chosen above this barrier, the hyper-spheres associated with the two inherent states 'merges' into a high dimensional peanut shape. We make the simplification that the model $U(\bf{R})=U_0$ surface is simply the surface of the union of the spheres. This means that special care has to be taken in performing the geodetic flow when reaching an intersection between hyper-spheres. This will be discussed below.

In general the hyper-spheres will have different sizes and be correlated in space. We make the simplifications that the hyper-spheres all have the same radius, $R$ (equal to  unity in the applied units), and they are randomly distributed in a hyper-cube with periodic boundary conditions.

To summarize the definition of the model  $U({\bf R})=U_0$ surface:  Let $K$ be a set of $H$ uniformly random distributed points inside the hyper-cube of size $L^n$ and $B_{\vec{p}}^{n}$ an hyper-sphere in $n$ dimensions centered at $p$. Then the model hyper-surface in $n$ dimensions is a $n$-dimensional surface, $M$, such that
\begin{equation}
M=\partial \left( \bigcup_{\vec{p} \in K} B_{\vec{p}}^{n} \right)
\end{equation}

Given the simplifications made, the model has only a single physically relevant parameter, the density of the hyper-spheres. We follow the literature on percolation \cite{Torquato2012a, Torquato2012b} and quantify this by the \emph{reduced density} $\eta$, i.e., the total volume of the hyper-spheres (ignoring overlap) divided by the total accessible volume: 
\begin{equation}
    \eta \equiv \frac{HV_{sphere}}{V_{cube}} = \frac{\pi^{\frac{n}{2}}}{\frac{n}{2}!} \frac{HR^n}{L^n}
    \label{eq:density}
\end{equation}

In high dimensions, the percolation density is given by $\eta_c = 2^{-n}$. Since we are looking for diffusive dynamics, we will focus on densities well above the percolation density.

\section{Geodetic flow}\label{sec:geodetic}

The hyper-surface composed of the union of intersecting hyper-sphere's surfaces is a differentiable manifold 
and consequently the geodetic flow is well defined. Examples of the geodetic flow on hyper-surfaces is given in Fig. \ref{fig:themodel3d}, in 3 and 16 dimensions respectively.
When the system is at the intersection between two spheres, it moves from the geodetic on the sphere it came from 
to a geodetic on the new sphere. The new geodetic is chosen in order to preserve the direction. 
This choice makes the transition from one sphere to the next one unique and reversible 
(details are given in the Supplementary Material).

In order to simulate the geodetic flow a neighbor list is created for the random distribution of hyper-spheres in the simulation box. Since the hyper-spheres are fixed, the calculation of a neighborlist is only needed at the start of the simulation.
When the system moves on an hyper-sphere,  it will follow a great circle until it meets an intersection with another hyper-sphere. The motion on a single sphere and where the particle meets an intersection can be calculated analytically 
(as discussed in the Supplementary Material).
Consequently, the motion of the particle can be evolved from an intersection to the next one in just one step. 
In this sense the algorithm used for simulating this system is similar to the event-driven dynamics used 
for hard spheres simulations.

\section{Results}

\begin{figure}[]
	\centering
	\resizebox{0.5\textwidth}{!}{
		\begin{tabular}{l}
			\begin{lpic}[]{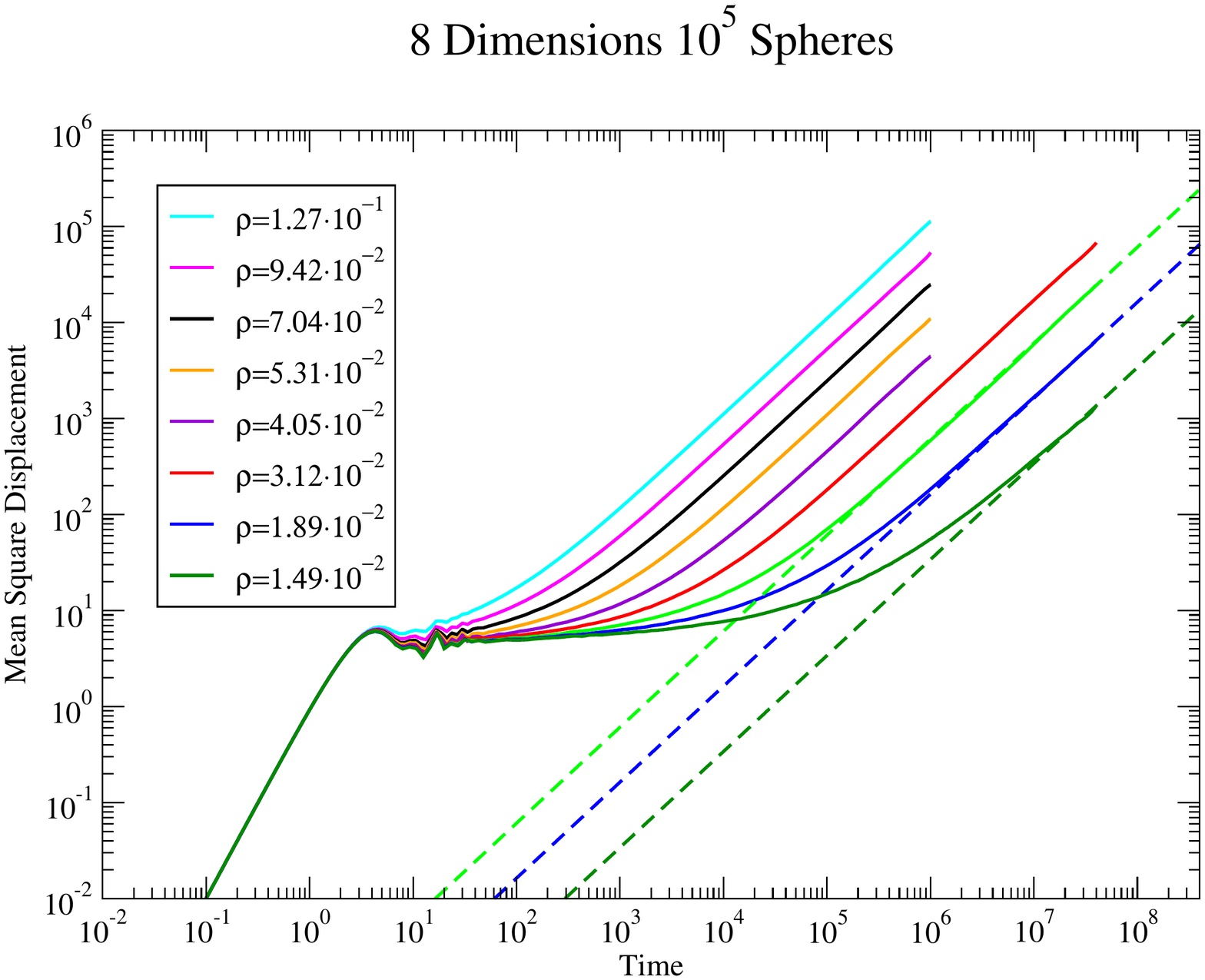}\end{lpic} \\
			\begin{lpic}[]{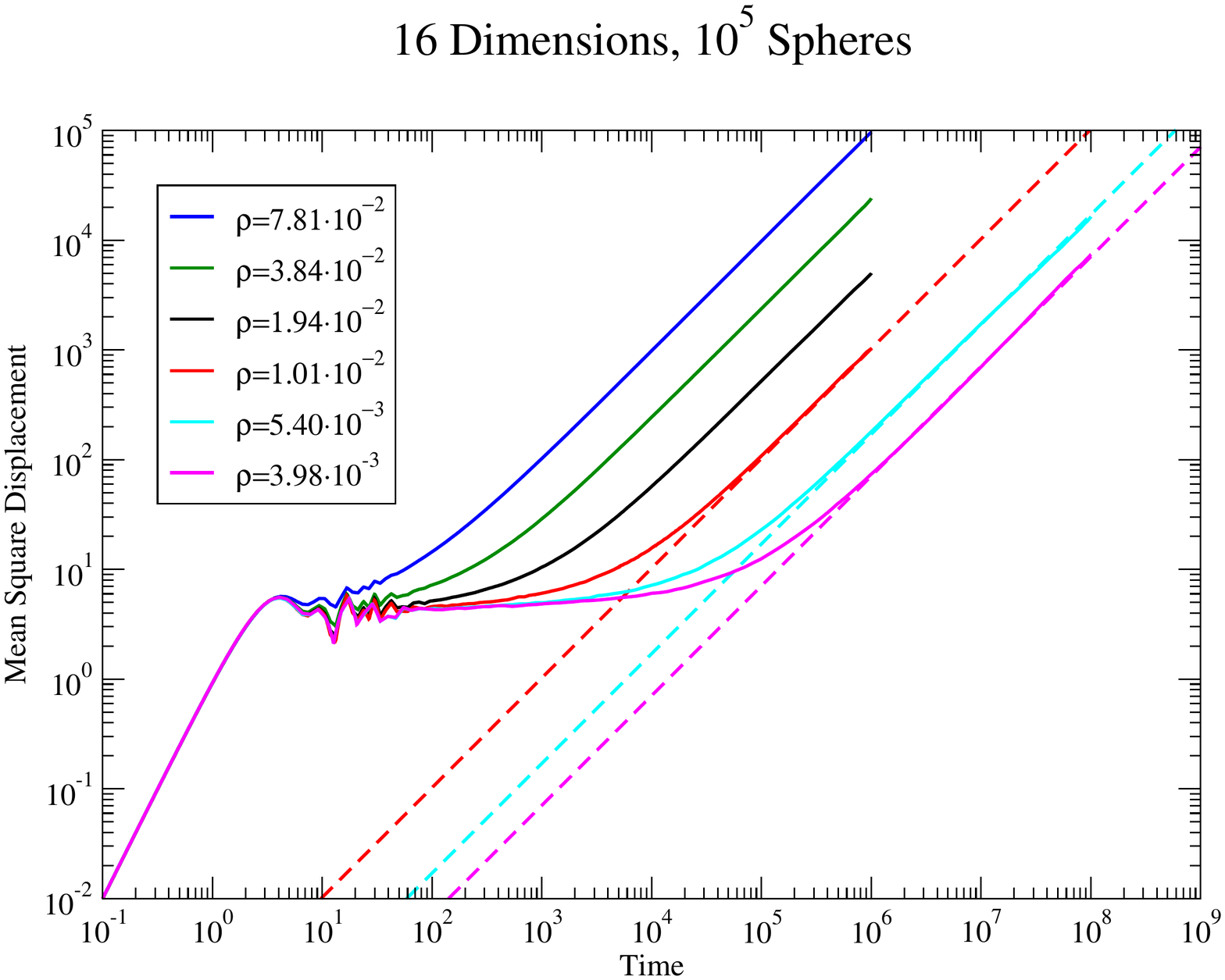}\lbl[]{230,50;\Huge{(a)}}\end{lpic} \\
	\end{tabular}}
	\caption{\label{fig:slow}
		mean square displacement as a function of time (i.e. curve length) the model system simulated in $8$ and $16$ dimensions with $H=10^5$ spheres. Dashed lines are fits to $MSD = 2Dt$, where $D$ is the diffusion coefficient.
	}
\end{figure}

The first desirable feature for a model of a supercooled liquid is to exhibit slow dynamics.
With this it is meant that for a change of some given control parameter, in this case the reduced density, $\eta$, the dynamics of the system should change from a regular liquid dynamics to a supercooled dynamics.
In Fig.\ref{fig:slow} the mean-squared displacement (MSD) is shown 8 and 16 dimensions, for different reduced densities. The curves look very similar to those usually observed for viscous liquids: at short times there is a ballistic regime $\propto t^2$, at long time there is a diffusive regime $\propto t$, and in-between the two there is a plateau that increases in length as the dynamics slows down - in this case when the reduced density is decreased. 

\begin{figure}[]
	\centering
	\resizebox{0.45\textwidth}{!}{
		\begin{tabular}{l}
			\begin{lpic}[]{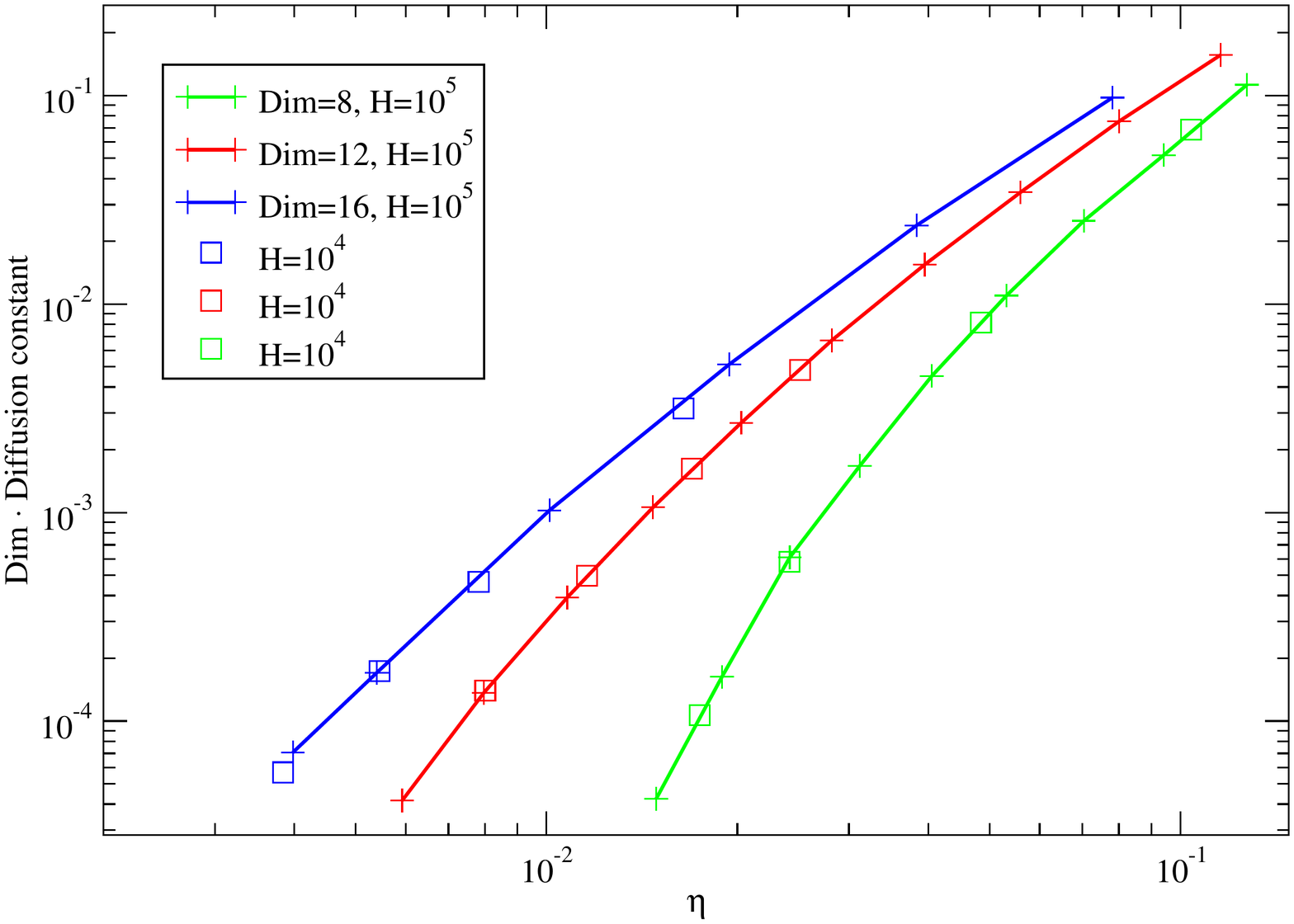}\end{lpic} \\
	\end{tabular}}
	\caption{Diffusion coefficient obtained from the MSD in Fig. \ref{fig:slow}, supplemented with results for 12 dimensions.
		Results for two system sizes are compared: $+$ for $10^5$ spheres and $\square$ for $10^4$ spheres.
		Full lines are guides for the eye.}
	\label{fig:diffvsrho}
\end{figure}

Fig. \ref{fig:diffvsrho} shows the diffusion coefficients as a function of reduced density for 8, 12, and 16 dimensions. Comparisons between $H=10^5$ spheres, as used in Fig.\ref{fig:slow}, and $H=10^4$ indicates that the samples used are large enough to avoid any significant finite-size effects.

\begin{figure}[]
\centering
  \resizebox{0.45\textwidth}{!}{
  \begin{tabular}{l}
  \begin{lpic}[]{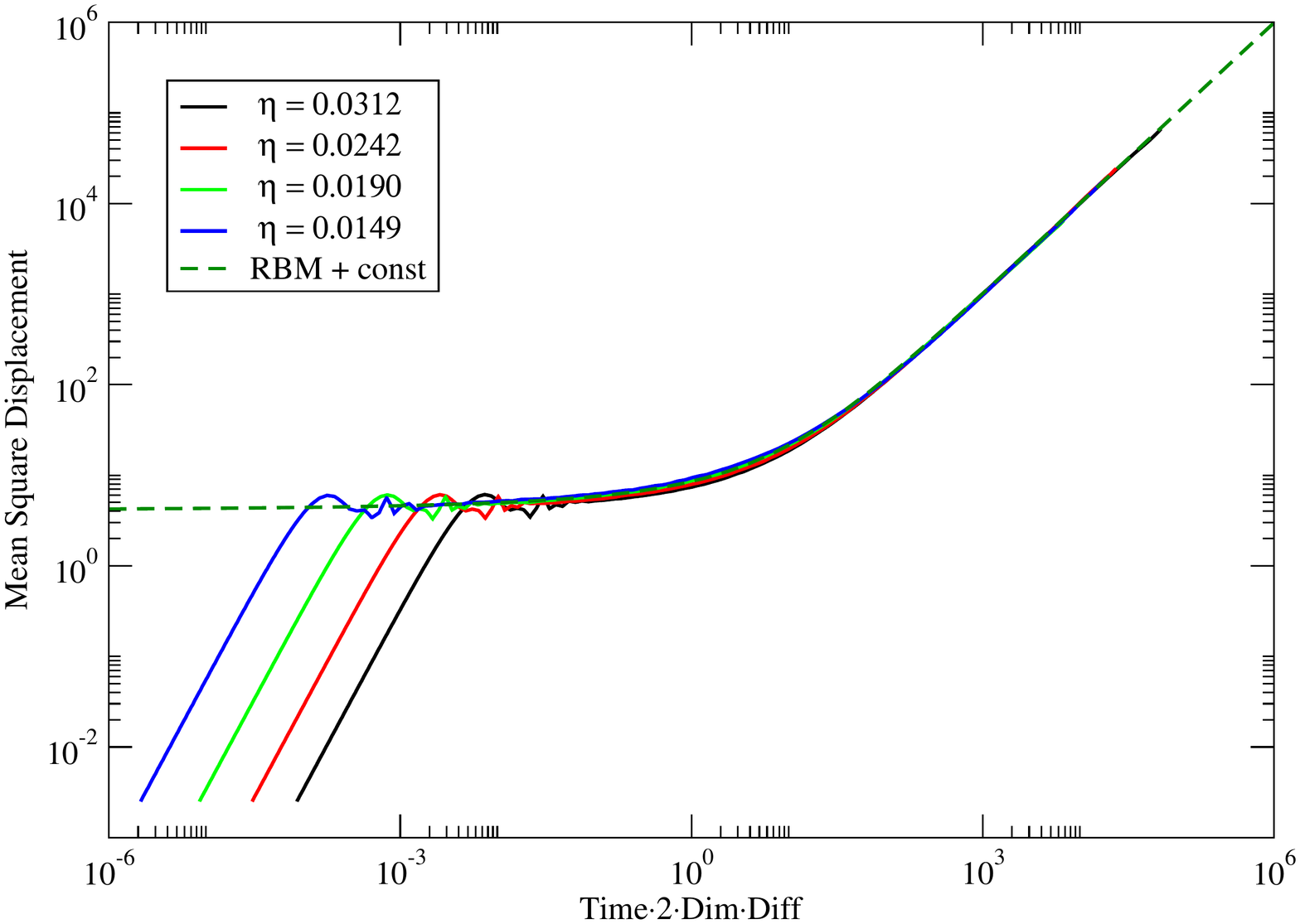}\end{lpic} \\
  \begin{lpic}[]{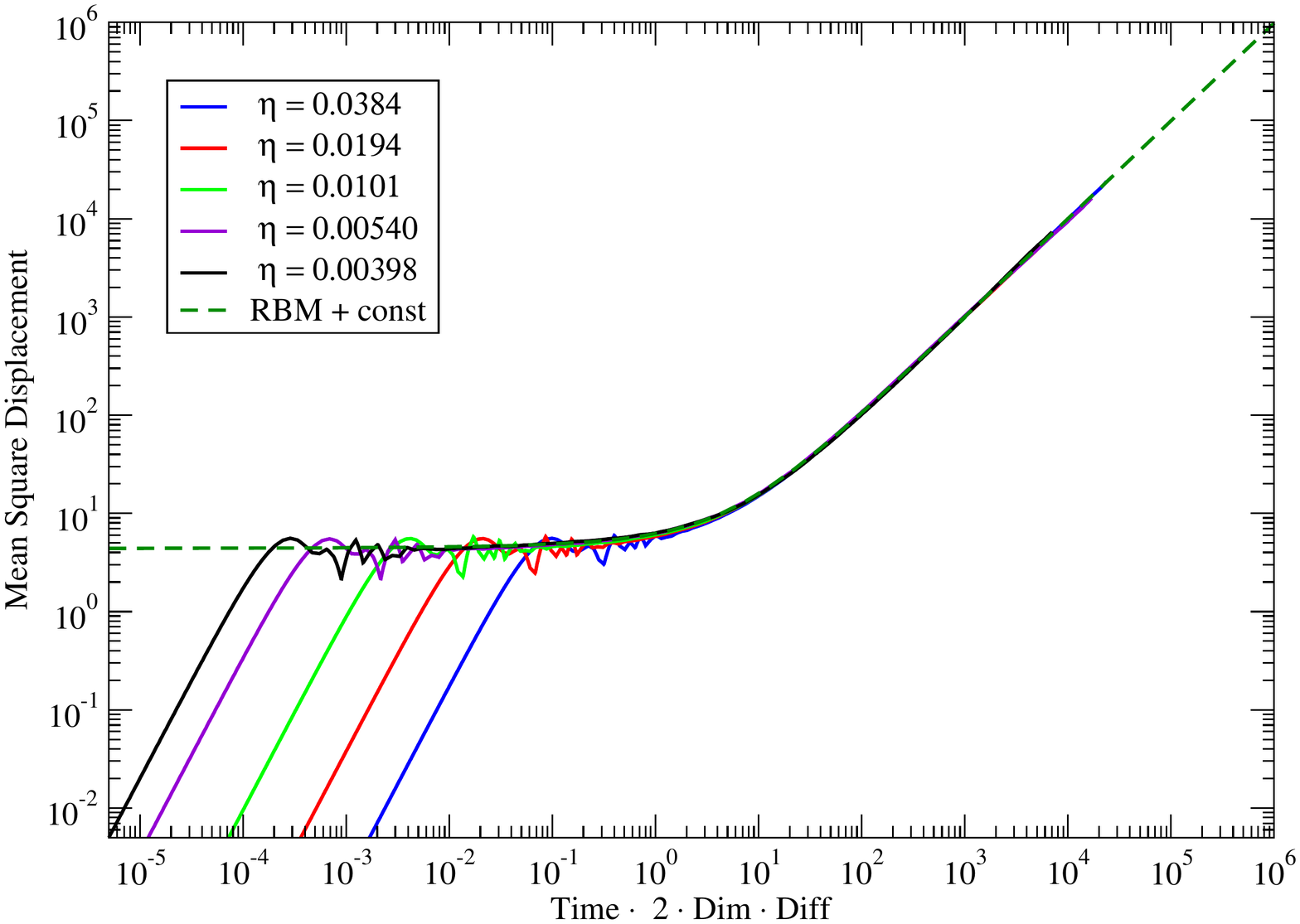}\end{lpic}
  \end{tabular}}
  \caption{Comparison between the shape of the slow dynamics MSD and the Random Barrier Model (RBM) \cite{RBM2020} for 8 and 16 dimensions.
  	The MSD's are plotted as a function of time scaled by the diffusion coefficients. The data collapse indicates time-temperature superposition (TTS). The prediction of the RMB is shown as dashed lines. 
  	A constant is added to the RBM result, reflecting the contribution from "cage rattling", which is not included in the RBM \cite{RBM2020}.
  }
  \label{fig:rbm}
\end{figure}

\begin{figure}[]
	\centering
	\resizebox{0.45\textwidth}{!}{
		\begin{tabular}{l}
			  \begin{lpic}[]{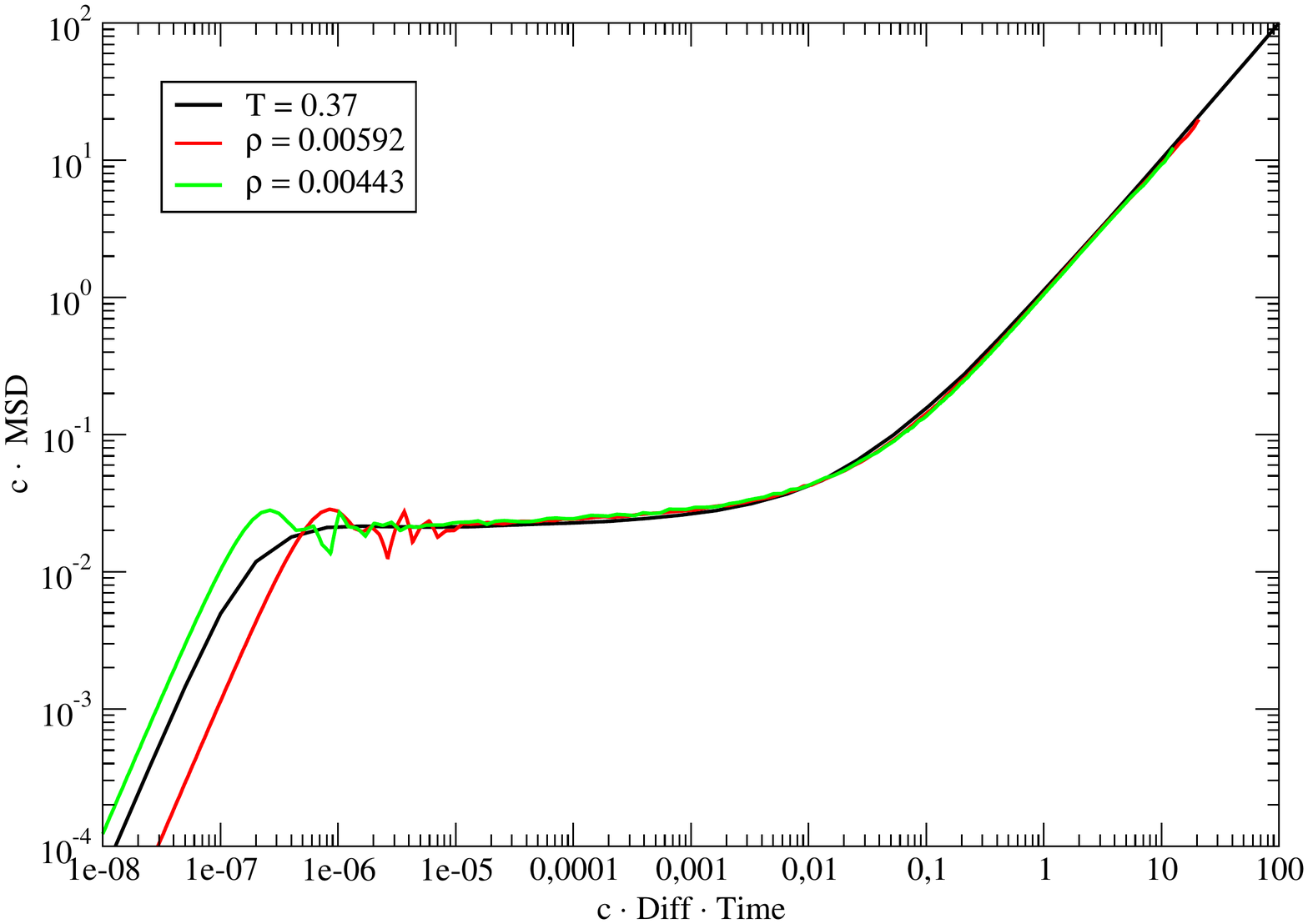}\end{lpic} \\
	\end{tabular}}
	\caption{Direct comparison between the shape of MSD for the KABLJ at T=0.37 (black curve, \cite{RBM2020}), and the new hyper-sphere model.}
	\label{fig:kablj}
\end{figure}

Often viscous liquids are found to exhibit time-temperature superposition (TTS), i.e., that the dynamic curves can be made to collapse by appropriate scaling. In Fig. \ref{fig:rbm} the mean-square displacements are plotted as a function of time scaled by the diffusion coefficient, for low reduced densities. Convincing data-collapse is observed, i.e. the model exhibits TTS.   

It has been recently shown that the shape of the MSD for the Kob and Andersen binary LJ mixture (KABLJ) system is well described in term of the Random Barrier Model \cite{RBM2020}. 
According to this model the energy landscape of disordered solids can be modeled as a cubic lattice whose sites correspond to potential energy minima, all with identical energies, separated by random energy barriers.
This model has been recently shown to provide a surprisingly accurate description of fluidity data in experiments \cite{Bierwirth2017}. 
In Fig. \ref{fig:rbm} the prediction of the RBM is included as dashed lines, showing very good agreement with the collapsed data for the mean-square displacement. We conclude, that the shape of the mean-square displacement curves are very similar to that of KABLJ. This conclusion is supported by Fig. \ref{fig:kablj} where the MSD of the new hyper-sphere model is compared directly with the MSD of a very low temperature state point for the KABLJ model.

\section{Conclusions}
In this manuscript a very simple model -- one might even say: a toy model -- for viscous liquid dynamics was proposed. In a single sentence the model is: "Geodetic flow on the surface of the union of randomly distributed hyper-spheres with unity radius". Despite its simplicity,the model was shown to reproduce several properties of the dynamics of viscous liquids: 
i) the MSD exhibits ballistic, plateau, and diffusion regimes
ii) drastic slowing down of the dynamics over several orders of magnitude;
iii) time-temperature superposition (TTS);
iv) the shape of the MSD curves are very well described by the RBM, and thus very close to the KABLJ.


Further work could be done studying the possible generalizations of this model which could be to consider 
hyper-spheres of different radius or to consider hyper-ellipsoids instead of hyper-spheres.


%
%

%


\let\cleardoublepage\clearpage
\bibliography{MyBibliografy}

\begin{thebibliography}{17}%
\makeatletter
\providecommand \@ifxundefined [1]{%
 \@ifx{#1\undefined}
}%
\providecommand \@ifnum [1]{%
 \ifnum #1\expandafter \@firstoftwo
 \else \expandafter \@secondoftwo
 \fi
}%
\providecommand \@ifx [1]{%
 \ifx #1\expandafter \@firstoftwo
 \else \expandafter \@secondoftwo
 \fi
}%
\providecommand \natexlab [1]{#1}%
\providecommand \enquote  [1]{``#1''}%
\providecommand \bibnamefont  [1]{#1}%
\providecommand \bibfnamefont [1]{#1}%
\providecommand \citenamefont [1]{#1}%
\providecommand \href@noop [0]{\@secondoftwo}%
\providecommand \href [0]{\begingroup \@sanitize@url \@href}%
\providecommand \@href[1]{\@@startlink{#1}\@@href}%
\providecommand \@@href[1]{\endgroup#1\@@endlink}%
\providecommand \@sanitize@url [0]{\catcode `\\12\catcode `\$12\catcode
  `\&12\catcode `\#12\catcode `\^12\catcode `\_12\catcode `\%12\relax}%
\providecommand \@@startlink[1]{}%
\providecommand \@@endlink[0]{}%
\providecommand \url  [0]{\begingroup\@sanitize@url \@url }%
\providecommand \@url [1]{\endgroup\@href {#1}{\urlprefix }}%
\providecommand \urlprefix  [0]{URL }%
\providecommand \Eprint [0]{\href }%
\providecommand \doibase [0]{https://doi.org/}%
\providecommand \selectlanguage [0]{\@gobble}%
\providecommand \bibinfo  [0]{\@secondoftwo}%
\providecommand \bibfield  [0]{\@secondoftwo}%
\providecommand \translation [1]{[#1]}%
\providecommand \BibitemOpen [0]{}%
\providecommand \bibitemStop [0]{}%
\providecommand \bibitemNoStop [0]{.\EOS\space}%
\providecommand \EOS [0]{\spacefactor3000\relax}%
\providecommand \BibitemShut  [1]{\csname bibitem#1\endcsname}%
\let\auto@bib@innerbib\@empty
\bibitem [{\citenamefont {Carnahan}\ and\ \citenamefont
  {Starling}(1969)}]{Carnahan1969}%
  \BibitemOpen
  \bibfield  {author} {\bibinfo {author} {\bibfnamefont {N.~F.}\ \bibnamefont
  {Carnahan}}\ and\ \bibinfo {author} {\bibfnamefont {K.~E.}\ \bibnamefont
  {Starling}},\ }\bibfield  {title} {\bibinfo {title} {Equation of state for
  nonattracting rigid spheres},\ }\href {https://doi.org/10.1063/1.1672048}
  {\bibfield  {journal} {\bibinfo  {journal} {The Journal of Chemical Physics}\
  }\textbf {\bibinfo {volume} {51}},\ \bibinfo {pages} {635} (\bibinfo {year}
  {1969})}\BibitemShut {NoStop}%
\bibitem [{\citenamefont {{Hansen, J.-P. and McDonald,
  I.R.}}(2006)}]{HansenMcDonald}%
  \BibitemOpen
  \bibfield  {author} {\bibinfo {author} {\bibnamefont {{Hansen, J.-P. and
  McDonald, I.R.}}},\ }\href@noop {} {\emph {\bibinfo {title} {{Theory of
  Simple Liquids}}}},\ \bibinfo {edition} {3rd}\ ed.\ (\bibinfo  {publisher}
  {Accademic},\ \bibinfo {address} {New York},\ \bibinfo {year}
  {2006})\BibitemShut {NoStop}%
\bibitem [{\citenamefont {Charbonneau}\ \emph {et~al.}(2017)\citenamefont
  {Charbonneau}, \citenamefont {Kurchan}, \citenamefont {Parisi}, \citenamefont
  {Urbani},\ and\ \citenamefont {Zamponi}}]{Charbonneau2017}%
  \BibitemOpen
  \bibfield  {author} {\bibinfo {author} {\bibfnamefont {P.}~\bibnamefont
  {Charbonneau}}, \bibinfo {author} {\bibfnamefont {J.}~\bibnamefont
  {Kurchan}}, \bibinfo {author} {\bibfnamefont {G.}~\bibnamefont {Parisi}},
  \bibinfo {author} {\bibfnamefont {P.}~\bibnamefont {Urbani}},\ and\ \bibinfo
  {author} {\bibfnamefont {F.}~\bibnamefont {Zamponi}},\ }\bibfield  {title}
  {\bibinfo {title} {Glass and jamming transitions: From exact results to
  finite-dimensional descriptions},\ }\href
  {https://doi.org/10.1146/annurev-conmatphys-031016-025334} {\bibfield
  {journal} {\bibinfo  {journal} {Annual Review of Condensed Matter Physics}\
  }\textbf {\bibinfo {volume} {8}},\ \bibinfo {pages} {265} (\bibinfo {year}
  {2017})}\BibitemShut {NoStop}%
\bibitem [{\citenamefont {Berthier}\ \emph {et~al.}(2016)\citenamefont
  {Berthier}, \citenamefont {Coslovich}, \citenamefont {Ninarello},\ and\
  \citenamefont {Ozawa}}]{SwapMC}%
  \BibitemOpen
  \bibfield  {author} {\bibinfo {author} {\bibfnamefont {L.}~\bibnamefont
  {Berthier}}, \bibinfo {author} {\bibfnamefont {D.}~\bibnamefont {Coslovich}},
  \bibinfo {author} {\bibfnamefont {A.}~\bibnamefont {Ninarello}},\ and\
  \bibinfo {author} {\bibfnamefont {M.}~\bibnamefont {Ozawa}},\ }\bibfield
  {title} {\bibinfo {title} {Equilibrium sampling of hard spheres up to the
  jamming density and beyond},\ }\href
  {https://doi.org/10.1103/PhysRevLett.116.238002} {\bibfield  {journal}
  {\bibinfo  {journal} {Phys. Rev. Lett.}\ }\textbf {\bibinfo {volume} {116}},\
  \bibinfo {pages} {238002} (\bibinfo {year} {2016})}\BibitemShut {NoStop}%
\bibitem [{\citenamefont {Gnan}\ and\ \citenamefont
  {Zaccarelli}(2019)}]{Gnan2019}%
  \BibitemOpen
  \bibfield  {author} {\bibinfo {author} {\bibfnamefont {N.}~\bibnamefont
  {Gnan}}\ and\ \bibinfo {author} {\bibfnamefont {E.}~\bibnamefont
  {Zaccarelli}},\ }\bibfield  {title} {\bibinfo {title} {The microscopic role
  of deformation in the dynamics of soft colloids},\ }\href
  {https://doi.org/10.1038/s41567-019-0480-1} {\bibfield  {journal} {\bibinfo
  {journal} {Nature Physics}\ }\textbf {\bibinfo {volume} {15}},\ \bibinfo
  {pages} {683} (\bibinfo {year} {2019})}\BibitemShut {NoStop}%
\bibitem [{\citenamefont {Stillinger}\ and\ \citenamefont
  {Weber}(1984)}]{Stillinger1984}%
  \BibitemOpen
  \bibfield  {author} {\bibinfo {author} {\bibfnamefont {F.~H.}\ \bibnamefont
  {Stillinger}}\ and\ \bibinfo {author} {\bibfnamefont {T.~A.}\ \bibnamefont
  {Weber}},\ }\bibfield  {title} {\bibinfo {title} {Packing structures and
  transitions in liquids and solids},\ }\href
  {https://doi.org/10.1126/science.225.4666.983} {\bibfield  {journal}
  {\bibinfo  {journal} {Science}\ }\textbf {\bibinfo {volume} {225}},\ \bibinfo
  {pages} {983} (\bibinfo {year} {1984})}\BibitemShut {NoStop}%
\bibitem [{\citenamefont {Debenedetti}\ \emph {et~al.}(1999)\citenamefont
  {Debenedetti}, \citenamefont {Stillinger}, \citenamefont {Truskett},\ and\
  \citenamefont {Roberts}}]{Debenedetti1999}%
  \BibitemOpen
  \bibfield  {author} {\bibinfo {author} {\bibfnamefont {P.~G.}\ \bibnamefont
  {Debenedetti}}, \bibinfo {author} {\bibfnamefont {F.~H.}\ \bibnamefont
  {Stillinger}}, \bibinfo {author} {\bibfnamefont {T.~M.}\ \bibnamefont
  {Truskett}},\ and\ \bibinfo {author} {\bibfnamefont {C.~J.}\ \bibnamefont
  {Roberts}},\ }\bibfield  {title} {\bibinfo {title} {The equation of state of
  an energy landscape},\ }\href {https://doi.org/10.1021/jp991384m} {\bibfield
  {journal} {\bibinfo  {journal} {The Journal of Physical Chemistry B}\
  }\textbf {\bibinfo {volume} {103}},\ \bibinfo {pages} {7390} (\bibinfo {year}
  {1999})}\BibitemShut {NoStop}%
\bibitem [{\citenamefont {Pedersen}\ \emph {et~al.}(2006)\citenamefont
  {Pedersen}, \citenamefont {Hecksher}, \citenamefont {Dyre},\ and\
  \citenamefont {Schr{\o}der}}]{Pedersen2006}%
  \BibitemOpen
  \bibfield  {author} {\bibinfo {author} {\bibfnamefont {U.~R.}\ \bibnamefont
  {Pedersen}}, \bibinfo {author} {\bibfnamefont {T.}~\bibnamefont {Hecksher}},
  \bibinfo {author} {\bibfnamefont {J.~C.}\ \bibnamefont {Dyre}},\ and\
  \bibinfo {author} {\bibfnamefont {T.~B.}\ \bibnamefont {Schr{\o}der}},\
  }\bibfield  {title} {\bibinfo {title} {An energy landscape model for
  glass-forming liquids in three dimensions},\ }\href
  {https://doi.org/https://doi.org/10.1016/j.jnoncrysol.2006.03.127} {\bibfield
   {journal} {\bibinfo  {journal} {Journal of Non-Crystalline Solids}\ }\textbf
  {\bibinfo {volume} {352}},\ \bibinfo {pages} {5210 } (\bibinfo {year}
  {2006})},\ \bibinfo {note} {proceedings of the 5th International Discussion
  Meeting on Relaxations in Complex Systems}\BibitemShut {NoStop}%
\bibitem [{\citenamefont {Schrøder}\ \emph {et~al.}(2000)\citenamefont
  {Schrøder}, \citenamefont {Sastry}, \citenamefont {Dyre},\ and\
  \citenamefont {Glotzer}}]{Schroder2000}%
  \BibitemOpen
  \bibfield  {author} {\bibinfo {author} {\bibfnamefont {T.~B.}\ \bibnamefont
  {Schrøder}}, \bibinfo {author} {\bibfnamefont {S.}~\bibnamefont {Sastry}},
  \bibinfo {author} {\bibfnamefont {J.~C.}\ \bibnamefont {Dyre}},\ and\
  \bibinfo {author} {\bibfnamefont {S.~C.}\ \bibnamefont {Glotzer}},\
  }\bibfield  {title} {\bibinfo {title} {Crossover to potential energy
  landscape dominated dynamics in a model glass-forming liquid},\ }\href
  {https://doi.org/10.1063/1.481621} {\bibfield  {journal} {\bibinfo  {journal}
  {The Journal of Chemical Physics}\ }\textbf {\bibinfo {volume} {112}},\
  \bibinfo {pages} {9834} (\bibinfo {year} {2000})}\BibitemShut {NoStop}%
\bibitem [{\citenamefont {Ingebrigtsen}\ \emph
  {et~al.}(2011{\natexlab{a}})\citenamefont {Ingebrigtsen}, \citenamefont
  {Toxvaerd}, \citenamefont {Heilmann}, \citenamefont {Schr{\o}der},\ and\
  \citenamefont {Dyre}}]{NVU1}%
  \BibitemOpen
  \bibfield  {author} {\bibinfo {author} {\bibfnamefont {T.~S.}\ \bibnamefont
  {Ingebrigtsen}}, \bibinfo {author} {\bibfnamefont {S.}~\bibnamefont
  {Toxvaerd}}, \bibinfo {author} {\bibfnamefont {O.~J.}\ \bibnamefont
  {Heilmann}}, \bibinfo {author} {\bibfnamefont {T.~B.}\ \bibnamefont
  {Schr{\o}der}},\ and\ \bibinfo {author} {\bibfnamefont {J.~C.}\ \bibnamefont
  {Dyre}},\ }\bibfield  {title} {\bibinfo {title} {Nvu dynamics. i. geodesic
  motion on the constant-potential-energy hypersurface},\ }\href
  {https://doi.org/10.1063/1.3623585} {\bibfield  {journal} {\bibinfo
  {journal} {The Journal of Chemical Physics}\ }\textbf {\bibinfo {volume}
  {135}},\ \bibinfo {pages} {104101} (\bibinfo {year}
  {2011}{\natexlab{a}})}\BibitemShut {NoStop}%
\bibitem [{\citenamefont {Ingebrigtsen}\ \emph
  {et~al.}(2011{\natexlab{b}})\citenamefont {Ingebrigtsen}, \citenamefont
  {Toxvaerd}, \citenamefont {Schr{\o}der},\ and\ \citenamefont {Dyre}}]{NVU2}%
  \BibitemOpen
  \bibfield  {author} {\bibinfo {author} {\bibfnamefont {T.~S.}\ \bibnamefont
  {Ingebrigtsen}}, \bibinfo {author} {\bibfnamefont {S.}~\bibnamefont
  {Toxvaerd}}, \bibinfo {author} {\bibfnamefont {T.~B.}\ \bibnamefont
  {Schr{\o}der}},\ and\ \bibinfo {author} {\bibfnamefont {J.~C.}\ \bibnamefont
  {Dyre}},\ }\bibfield  {title} {\bibinfo {title} {Nvu dynamics. ii. comparing
  to four other dynamics},\ }\href {https://doi.org/10.1063/1.3623586}
  {\bibfield  {journal} {\bibinfo  {journal} {The Journal of Chemical Physics}\
  }\textbf {\bibinfo {volume} {135}},\ \bibinfo {pages} {104102} (\bibinfo
  {year} {2011}{\natexlab{b}})}\BibitemShut {NoStop}%
\bibitem [{\citenamefont {{Jeppe C. Dyre}}(2013)}]{quasiuniNVU}%
  \BibitemOpen
  \bibfield  {author} {\bibinfo {author} {\bibnamefont {{Jeppe C. Dyre}}},\
  }\bibfield  {title} {\bibinfo {title} {{NVU perspective on simple liquids'
  quasiuniversality}},\ }\href@noop {} {\bibfield  {journal} {\bibinfo
  {journal} {Physical Review E}\ }\textbf {\bibinfo {volume} {87}},\ \bibinfo
  {pages} {022106} (\bibinfo {year} {2013})}\BibitemShut {NoStop}%
\bibitem [{\citenamefont {Goldstein}(1969)}]{Goldstein1969}%
  \BibitemOpen
  \bibfield  {author} {\bibinfo {author} {\bibfnamefont {M.}~\bibnamefont
  {Goldstein}},\ }\bibfield  {title} {\bibinfo {title} {Viscous liquids and the
  glass transition: A potential energy barrier picture},\ }\href
  {https://doi.org/10.1063/1.1672587} {\bibfield  {journal} {\bibinfo
  {journal} {The Journal of Chemical Physics}\ }\textbf {\bibinfo {volume}
  {51}},\ \bibinfo {pages} {3728} (\bibinfo {year} {1969})}\BibitemShut
  {NoStop}%
\bibitem [{\citenamefont {Torquato}(2012)}]{Torquato2012a}%
  \BibitemOpen
  \bibfield  {author} {\bibinfo {author} {\bibfnamefont {S.}~\bibnamefont
  {Torquato}},\ }\bibfield  {title} {\bibinfo {title} {Effect of dimensionality
  on the continuum percolation of overlapping hyperspheres and hypercubes},\
  }\href {https://doi.org/10.1063/1.3679861} {\bibfield  {journal} {\bibinfo
  {journal} {The Journal of Chemical Physics}\ }\textbf {\bibinfo {volume}
  {136}},\ \bibinfo {pages} {054106} (\bibinfo {year} {2012})}\BibitemShut
  {NoStop}%
\bibitem [{\citenamefont {Torquato}\ and\ \citenamefont
  {Jiao}(2012)}]{Torquato2012b}%
  \BibitemOpen
  \bibfield  {author} {\bibinfo {author} {\bibfnamefont {S.}~\bibnamefont
  {Torquato}}\ and\ \bibinfo {author} {\bibfnamefont {Y.}~\bibnamefont
  {Jiao}},\ }\bibfield  {title} {\bibinfo {title} {Effect of dimensionality on
  the continuum percolation of overlapping hyperspheres and hypercubes. ii.
  simulation results a nd analyses},\ }\href
  {https://doi.org/10.1063/1.4742750} {\bibfield  {journal} {\bibinfo
  {journal} {The Journal of Chemical Physics}\ }\textbf {\bibinfo {volume}
  {137}},\ \bibinfo {pages} {074106} (\bibinfo {year} {2012})}\BibitemShut
  {NoStop}%
\bibitem [{\citenamefont {Schrøder}\ and\ \citenamefont
  {Dyre}(2020)}]{RBM2020}%
  \BibitemOpen
  \bibfield  {author} {\bibinfo {author} {\bibfnamefont {T.~B.}\ \bibnamefont
  {Schrøder}}\ and\ \bibinfo {author} {\bibfnamefont {J.~C.}\ \bibnamefont
  {Dyre}},\ }\bibfield  {title} {\bibinfo {title} {Solid-like mean-square
  displacement in glass-forming liquids},\ }\href
  {https://doi.org/10.1063/5.0004093} {\bibfield  {journal} {\bibinfo
  {journal} {The Journal of Chemical Physics}\ }\textbf {\bibinfo {volume}
  {152}},\ \bibinfo {pages} {141101} (\bibinfo {year} {2020})}\BibitemShut
  {NoStop}%
\bibitem [{\citenamefont {Bierwirth}\ \emph {et~al.}(2017)\citenamefont
  {Bierwirth}, \citenamefont {B\"ohmer},\ and\ \citenamefont
  {Gainaru}}]{Bierwirth2017}%
  \BibitemOpen
  \bibfield  {author} {\bibinfo {author} {\bibfnamefont {S.~P.}\ \bibnamefont
  {Bierwirth}}, \bibinfo {author} {\bibfnamefont {R.}~\bibnamefont
  {B\"ohmer}},\ and\ \bibinfo {author} {\bibfnamefont {C.}~\bibnamefont
  {Gainaru}},\ }\bibfield  {title} {\bibinfo {title} {Generic primary
  mechanical response of viscous liquids},\ }\href
  {https://doi.org/10.1103/PhysRevLett.119.248001} {\bibfield  {journal}
  {\bibinfo  {journal} {Phys. Rev. Lett.}\ }\textbf {\bibinfo {volume} {119}},\
  \bibinfo {pages} {248001} (\bibinfo {year} {2017})}\BibitemShut {NoStop}%
\end{thebibliography}%

\end{document}